\documentclass{moriond}

\def\btag{{B}_{\rm tag}}
\def\bsig{{B}_{\rm sig}}

\newcommand*{\RD}{\ensuremath{\mathcal{R}(D) }}
\newcommand*{\RDSt}{\ensuremath{\mathcal{R}(D^{\ast}) }}
\newcommand*{\RDall}{\ensuremath{\mathcal{R}(D^{(\ast)}) }}

\newcommand*{\eecl}{\ensuremath{E_{\rm ECL} }}
\newcommand*{\costheta}{\ensuremath{\cos\theta_{B,D^{(\ast)}\ell} }}

\newcommand*{\tauDec}{\ensuremath{\tau^- \to \ell^- \bar{\nu}_{\ell} \nu_{\tau} }}

\usepackage{graphicx}
\usepackage{dcolumn}
\usepackage{bm}
\usepackage{hyperref}
\usepackage[mathlines]{lineno}
\usepackage{multirow}
\usepackage{color}
\usepackage{url}
\usepackage{xcolor}
\usepackage{wrapfig}
\usepackage[skip=1pt]{caption}

\hypersetup{
}

\graphicspath{{imports/figures/}}

\newcommand{\Photo}{}

\begin{document}
\vspace*{4cm}
\title{Measurement of \RD\ and \RDSt\ with a semileptonic tagging method}

\author{G. Caria on behalf of the Belle Collaboration}

\address{School of Physics, University of Melbourne, Victoria 3010}

\setlength\intextsep{0pt}

\maketitle\abstracts{
We report a measurement of the ratios of branching fractions $\RDall = {\cal B}(\bar{B} \to D^{(*)} \tau^- \bar{\nu}_{\tau})/{\cal B}(\bar{B} \to D^{(*)} \ell^- \bar{\nu}_{\ell})$, where $\ell$ denotes an electron or a muon. The results are based on a data sample containing $772\times10^6$ $B\bar{B}$ events recorded at the $\Upsilon(4S)$ resonance with the Belle detector at the KEKB $e^+ e^-$ collider. The tag-side $B$ meson is reconstructed in a semileptonic decay mode, and the signal-side $\tau$ is reconstructed in a purely leptonic decay. The results are $\RD\ = 0.307 \pm 0.037 \pm 0.016$ and $\RDSt\ = 0.283 \pm 0.018 \pm 0.014$, where the first uncertainties are statistical and the second are systematic. These results are in agreement with the Standard Model predictions within $0.2$ and $1.1$ standard deviations, respectively.}


\section{Introduction}
Semitauonic $B$ meson decays of the type $b \to c \tau \nu_{\tau}$~\cite{note} are sensitive probes for physics beyond the Standard Model (SM). The ratio of branching fractions $\RDall = {\cal B}(\bar{B} \to D^{(*)} \tau^- \bar{\nu}_{\tau})/{\cal B}(\bar{B} \to D^{(*)} \ell^- \bar{\nu}_{\ell})$, where $\ell$ denotes an electron or a muon, is typically measured instead of the absolute branching fraction of $\bar{B} \to D^{(*)} \tau^- \bar{\nu}_{\tau}$ to reduce common systematic uncertainties. Hereafter, $\bar{B} \to D^{(*)} \tau^- \bar{\nu}_{\tau}$ and $\bar{B} \to D^{(*)} \ell^- \bar{\nu}_{\ell}$ will be referred to as the signal and normalization modes, respectively. The SM calculations for these ratios, performed by several groups~\cite{Bigi:2016mdz}\cite{Bernlochner:2017jka, Bigi:2017jbd, Jaiswal:2017rve}, are averaged~\cite{HFLAV} to obtain
$\RD  = 0.299 \pm 0.003$ and
$\RDSt    = 0.258 \pm 0.005$.
The average values of the experimental results are
$\RD = 0.407 \pm 0.039 \pm 0.024$ and
$\RDSt = 0.306 \pm 0.013 \pm 0.007$~\cite{HFLAV},
where the first uncertainty is statistical and the second is systematic. These values exceed SM predictions by $2.1\sigma$ and $3.0\sigma$, respectively. A combined analysis of \RD\ and \RDSt\, taking correlations into account, finds that the deviation from the SM prediction is approximately $3.8\sigma$~\cite{HFLAV}.

In this paper, we report the first measurement of \RD\ using the semileptonic tagging method, and update our earlier measurement of \RDSt~\cite{Sato:2016svk} by combining results of $B^0$ and $B^+$ decays using a more efficient tag reconstruction algorithm.
We use the full $\Upsilon(4S)$ data sample containing $772 \times 10^6$ $B \bar{B}$ events recorded with the Belle detector~\cite{Abashian:2000cg} at the KEKB $e^+ e^-$ collider~\cite{KUROKAWA20031}.
The Belle detector is described in detail elsewhere~\cite{Abashian:2000cg}. To determine the reconstruction efficiency and probability density functions (PDFs)  for signal, normalization, and background modes, we use Monte Carlo (MC) simulated events, generated with the EvtGen event generator~\cite{LANGE2001152}, and having the detector response simulated with the GEANT3 package~\cite{Brun:1987ma}.


\section{Event Reconstruction And Selection}
The $\btag$ is reconstructed using a hierarchical algorithm based on ``Fast" boosted decision trees (BDT)~\cite{Keck2019} in the $D^{} \ell \bar{\nu}_{\ell}$ and $D^{*} \ell \bar{\nu}_{\ell}$ channels, where $\ell = e, \mu$. We select well-reconstructed $\btag$ candidates by requiring their classifier output to be larger than $10^{-1.5}$. We veto ${B} \to D^{\ast} \tau (\to \ell \nu \nu) \nu  $ events on the tag side by applying a selection on  \costheta, defined as
$ \costheta \equiv 
\frac{2E_{\rm beam} E_{D^{(*)} \ell} - m_B^2 - m_{D^{(*)} \ell}^2}
{2 |\bm{p}_B| |\bm{p}_{D^{(*)} \ell}|}
$
where $E_{\rm beam}$ is the beam energy, and $E_{D^* \ell}$, $\bm{p}_{D^* \ell}$, and $m_{D^* \ell}$ are the energy, momentum, and mass, respectively, of the $D^* \ell$ system. The quantity $m_B$ is the nominal $B$ meson mass~\cite{PDG}, and $\bm{p}_B$ is the $B$ meson momentum. All quantities are evaluated in the $\Upsilon(4S)$ rest frame. Correctly reconstructed $B$ candidates in the normalization mode are expected to have a value of  \costheta\ between $-1$ and $+1$. Similarly, correctly reconstructed and misreconstructed $B$ candidates in the signal mode tend to have \costheta\ values more negative than this range due to additional missing particles. We account for detector resolution effects and apply the requirement $ -2.0 <  \costheta < 1.0$ for the $\btag$. 

In each event with a selected $\btag$ candidate, we search for the signature $D^{(*)} \ell$ among the remaining tracks and calorimeter clusters. We define four disjoint data samples, denoted $D^{+} \ell^{-}$, $D^{0} \ell^{-}$, $D^{*+} \ell^{-}$, and $D^{*0} \ell^{-}$.
On the signal side, neutral $D$ mesons are reconstructed in the following decay modes:
$D^0 \to K^- \pi^+ \pi^0$,
$K^- \pi^+ \pi^+ \pi^- $,
$K^- \pi^+$,
$K_S^0 \pi^+ \pi^-$,
$K_S^0 \pi^0$,
$K_S^0 K^+ K^-$,
$K^- K^+$ and
$\pi^- \pi^+$.
Similarly, charged $D$ mesons are reconstructed in the following modes:
$D^+ \to K^- \pi^+ \pi^+$,
$K_S^0 \pi^+ \pi^0$,
$K_S^0 \pi^+ \pi^+ \pi^-$,
$K_S^0 \pi^+$,
$K^- K^+ \pi^+$ and
$K_S^0 K^+$.
The combined branching fractions for reconstructed channels are 30\% and 22\% for $D^0$ and $D^+$, respectively. 
On the signal side, we require \costheta\ to be less than 1.0 and the $D^{(*)}$ momentum in the $\Upsilon(4S)$ rest frame to be less than 2.0 GeV/$c$.
Finally, we require that events contain no extra charged tracks, $K_S^0$ candidates, or $\pi^0$ candidates, which are reconstructed with the same criteria as those used for the $D$ candidates. When multiple $\btag$ or  $\bsig$ candidates are found in an event, we select the $\btag$ candidate with the highest tagging classifier output, and the $\bsig$ candidate with the highest p-value resulting from the $D$ or $D^*$ vertex fit. 

\section{Signal Extraction}

To distinguish signal and normalization events from background processes, we use the sum of the energies of neutral clusters detected in the ECL that are not associated with reconstructed particles, denoted as \eecl. We require that \eecl\ be less than 1.2 GeV.
To separate reconstructed signal and normalization events, we employ a BDT based on the \verb|XGBoost| package~\cite{xgboost16}. The input variables to the BDT are  \costheta; the approximate missing mass squared $m_{\rm miss}^2 = ( E_{\rm beam} - E_{D^{(*)}} - E_\ell )^2 - (\bm{p}_{D^{(*)}} + \bm{p}_{\ell} )^2$; the visible energy $E_{\rm vis} = \sum_i E_i$, where $(E_i, \bm{p}_i)$ is the four-momentum of particle $i$. We do not apply any selection on the BDT classifier output, denoted as \verb|class|; instead we use it as one of the fitting variables for the extraction of \RDall.

We extract the yields of signal and normalization modes from a two-dimensional (2D) extended maximum-likelihood fit to the variables \verb|class| and \eecl. The fit is performed simultaneously to the four $D^{(*)} \ell$ samples. 
The free parameters in the final fit are the yields of signal, normalization, $B \to D^{**} \ell \nu_{\ell}$ and feed-down from $D^* \ell$ to $D \ell$ components. The yields of other backgrounds are fixed to their MC expected values.
The ratios \RDall\ are given by the formula:
$ \RDall =
    \frac{1}{2{\cal B}(\tau^- \to \ell^- \bar{\nu}_{\ell} \nu_{\tau})}
    \cdot
    \frac{\varepsilon_{\rm norm}}{\varepsilon_{\rm sig}}
    \cdot
    \frac{N_{\rm sig}}{N_{\rm norm}}$
where $\varepsilon_{\rm sig (norm)}$ and $N_{\rm sig (norm)}$ are the detection efficiency and yields of signal (normalization) modes and ${\cal B}(\tau^- \to \ell^- \bar{\nu}_{\ell} \nu_{\tau})$ is the average of the world averages for $\ell =e$ and $\ell = \mu$.
To improve the accuracy of the MC simulation, we apply a series of correction factors determined from control sample measurements. Furthermore, we reweight events to account for differing yields of misreconstructed $D^{(*)}$ and for the different reconstruction efficiency of the tagging algorithm  between data and MC simulations.


\section{Systematic Uncertainties}

The systematic uncertainties are summarized in Table~\ref{tab:sys}, where the label ``$D^{**}$ composition" refers to the uncertainty introduced by the branching fractions of the $B \to D^{**} \ell \nu_{\ell}$ channels and the decays of the $D^{**}$ mesons, which are not well known and hence contribute significantly to the total PDF uncertainty due to $B \to D^{**} \ell \nu_{\ell}$ decays. A large systematic uncertainty arises from the limited size of MC samples, which contributes to the uncertainty in the PDF shapes and various efficiency factors used in the fit. The total systematic uncertainty is estimated by summing all contributions in quadrature.


\section{Results and Conclusion}

\begin{figure}[h]
    \begin{minipage}{0.50\linewidth}
    \centerline{\includegraphics[width=1\linewidth]{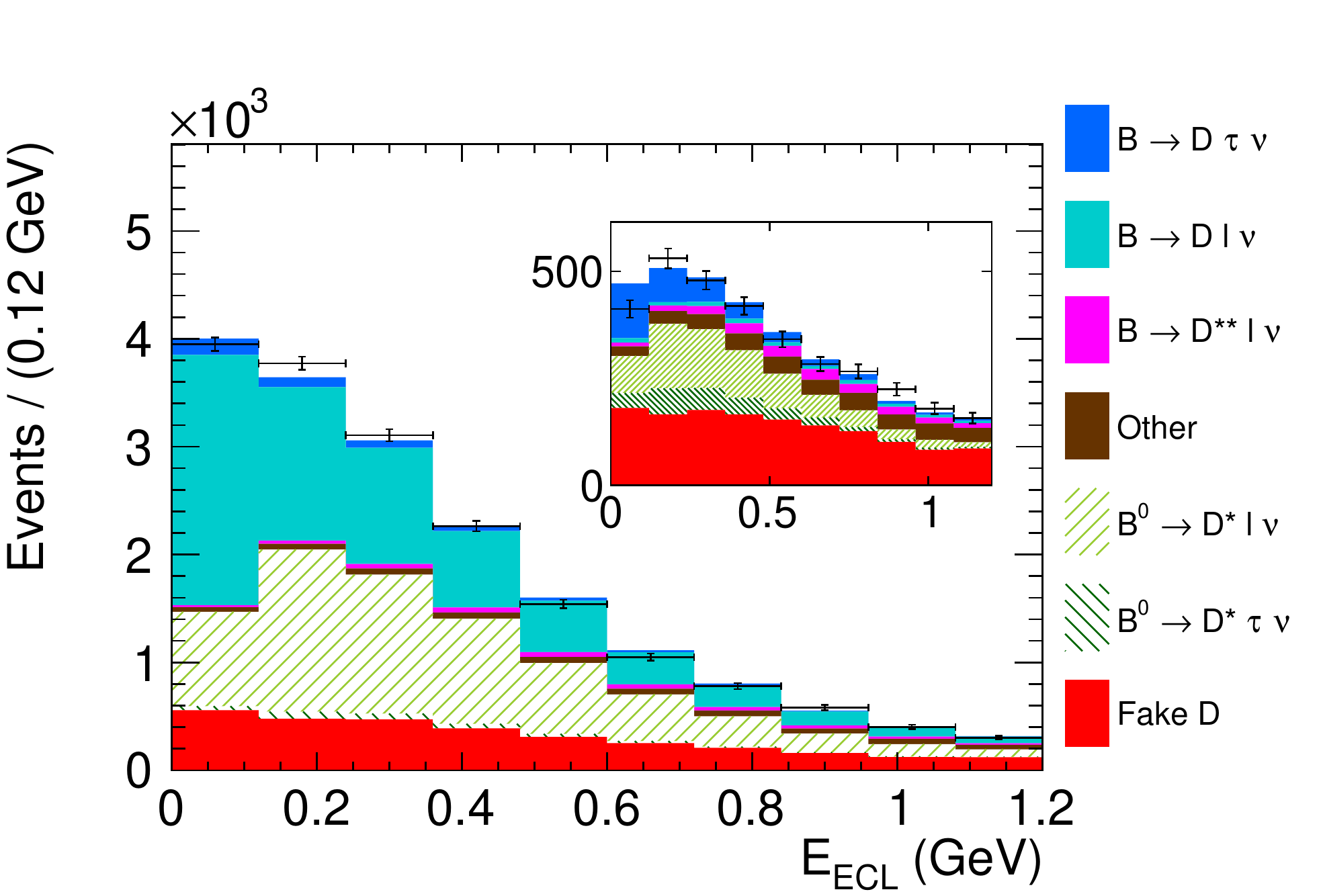}}
    \end{minipage}
    \hfill
    \begin{minipage}{0.50\linewidth}
    \centerline{\includegraphics[width=1\linewidth]{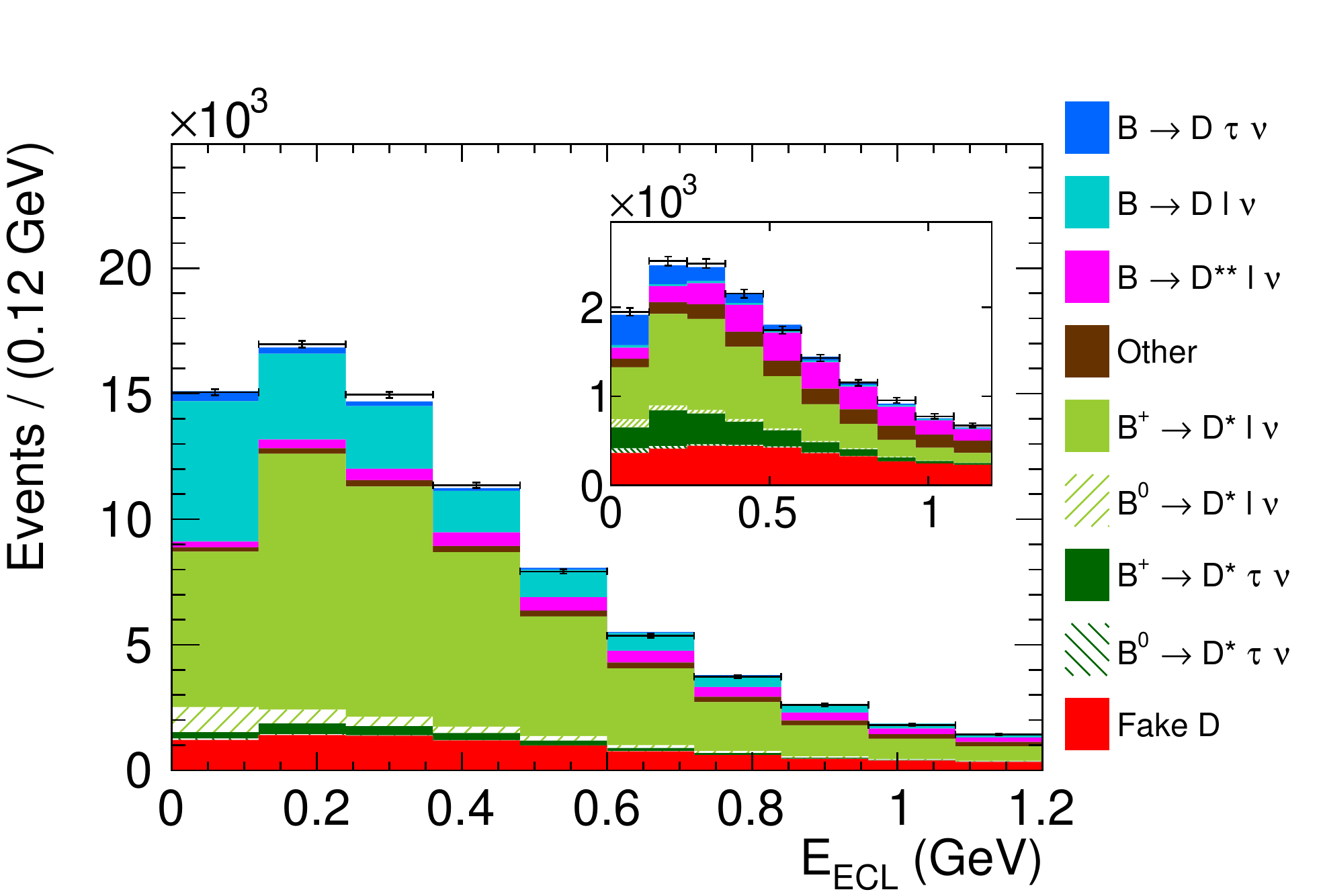}}
    \end{minipage}
    \hfill
    \begin{minipage}{0.50\linewidth}
    \centerline{\includegraphics[width=1\linewidth]{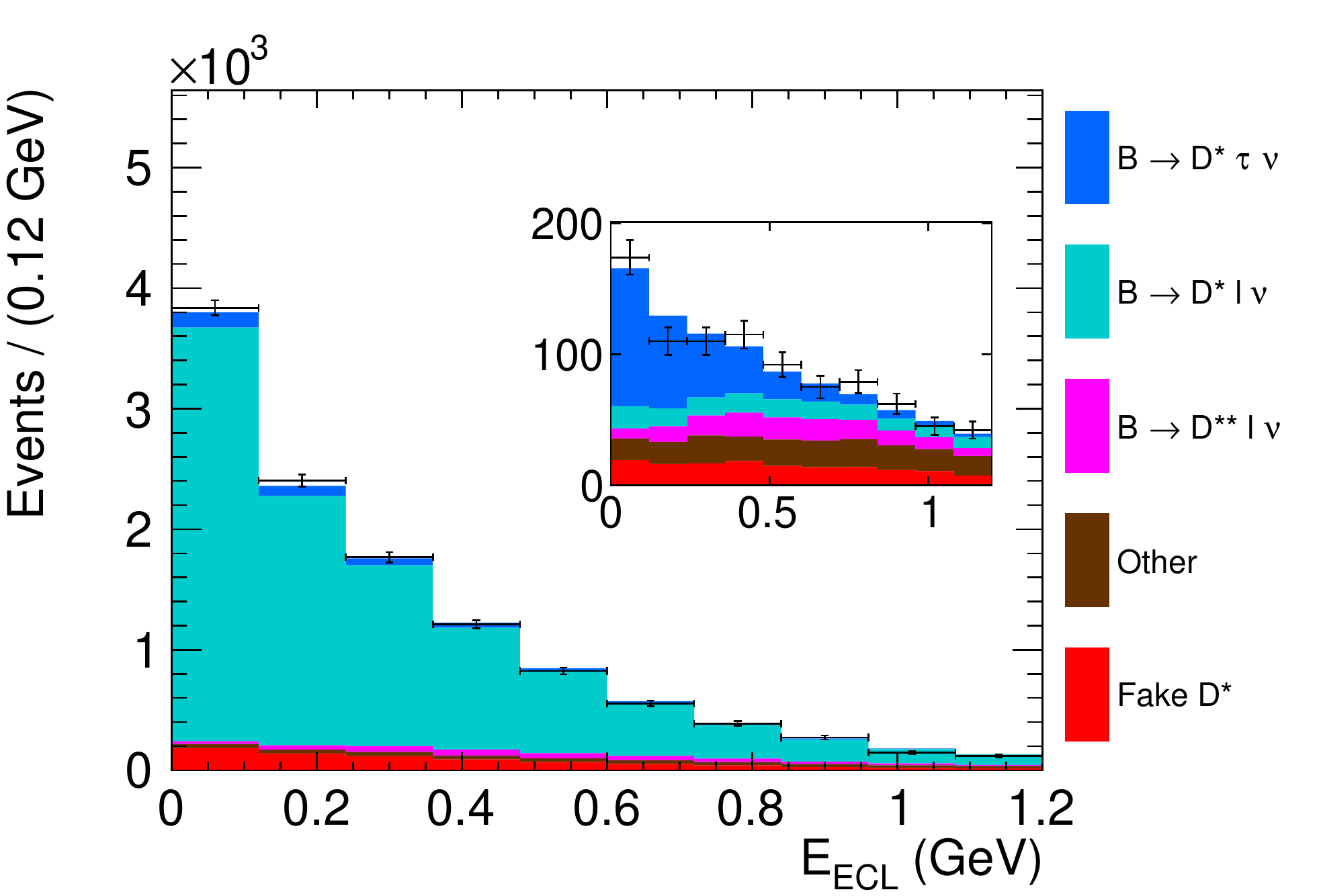}}
    \end{minipage}
    \hfill
    \begin{minipage}{0.50\linewidth}
    \centerline{\includegraphics[width=1\linewidth]{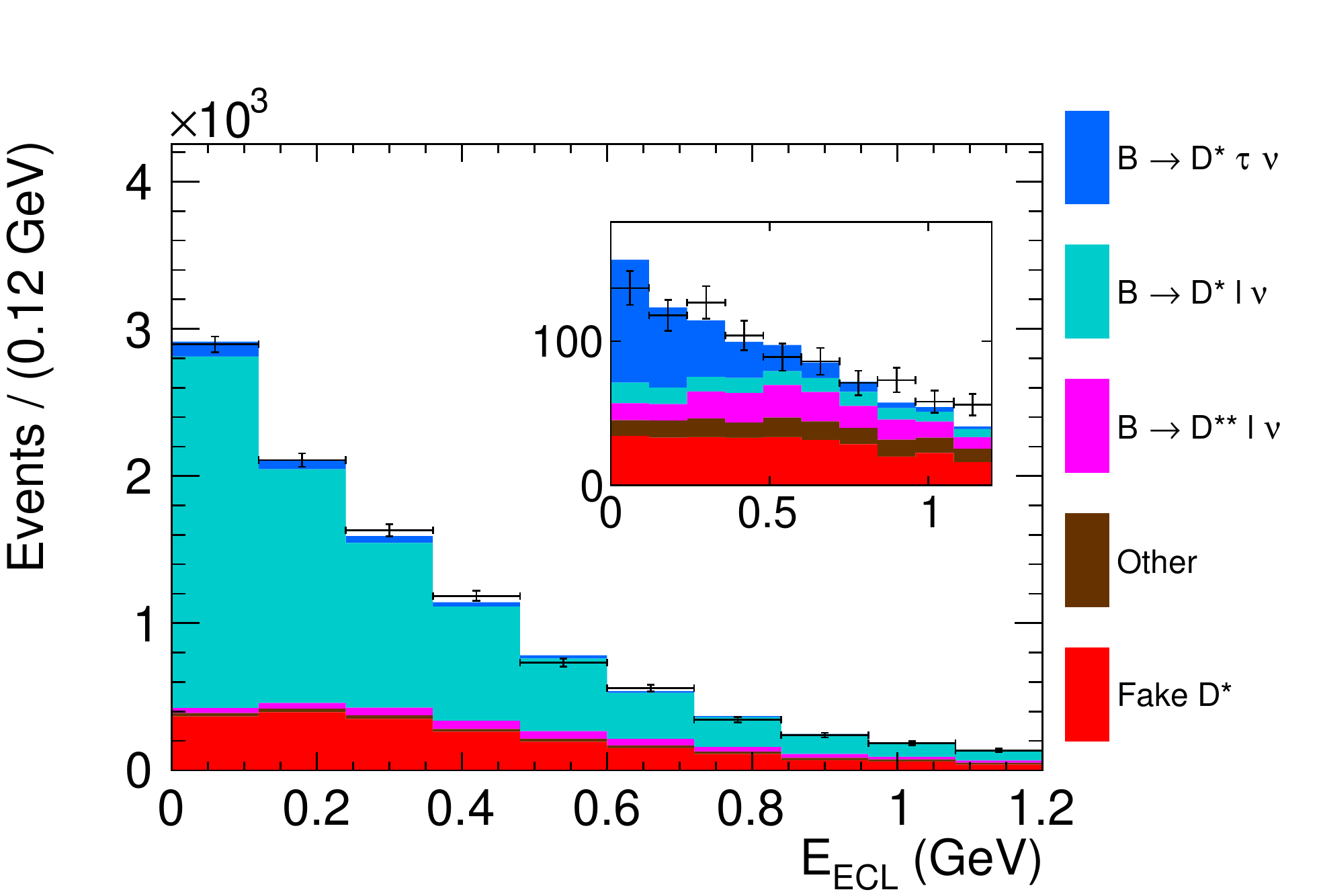}}
    \end{minipage}
    \hfill
    \caption{\eecl\ fit projections and data points with statistical uncertainties in the $D^+\ell^-$ (top left), $D^0\ell^-$ (top right), $D^{*+}\ell^-$ (bottom left) and $D^{*0}\ell^-$ (bottom right) samples, for the full classifier region. The signal region defined by the selection \texttt{class} $>0.9$ is shown in the inset.}
    \label{fig:results_allModes}
\end{figure}

Our results for the measurements of the ratios $\RDall = {\cal B}(\bar{B} \to D^{(*)} \tau^- \bar{\nu}_{\tau})/{\cal B}(\bar{B} \to D^{(*)} \ell^- \bar{\nu}_{\ell})$, where $\ell$ denotes an electron or a muon, based on a semileptonic tagging method using a data sample containing $772 \times 10^6 B\bar{B}$ events collected with the Belle detector, are
\begin{eqnarray}
	\RD 	&=& 0.307 \pm 0.037 \pm 0.016 \\
	\RDSt   &=& 0.283 \pm 0.018 \pm 0.014,
\end{eqnarray}
which are in agreement with the SM predictions within $0.2\sigma$ and $1.1\sigma$, respectively. The combined result agrees with the SM predictions within $1.2\sigma$. The \eecl\ projections of the fit are shown in Figure \ref{fig:results_allModes}. The 2D combination of the \RD\ and \RDSt\ results of this analysis are shown in Figure~\ref{fig:2Dplane}. This work constitutes the most precise measurements of \RD\ and \RDSt\ performed to date and the first result for \RD\ based on a semileptonic tagging method.

\begin{table}
\begin{minipage}{0.47\linewidth}
\centering
\includegraphics[width=\linewidth]{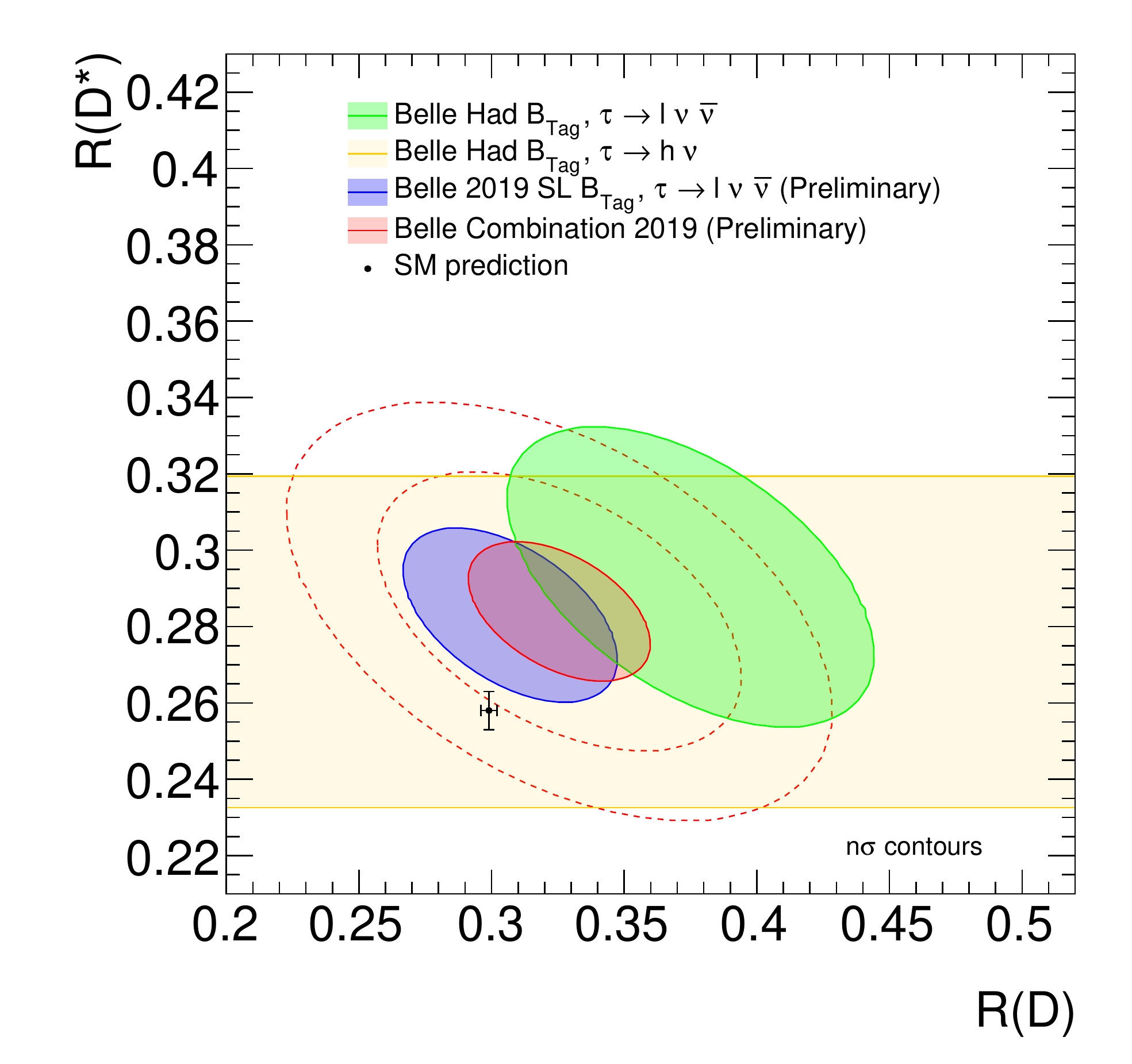}
\vspace{-0.8cm}
    \captionof{figure}{Fit results are shown on a \RD\ vs. \RDSt\ plane with the Belle results on \RD\ and \RDSt\ performed with a hadronic tag~\protect\cite{huschle2015, Hirose:2016wfn} and the preliminary Belle average.}
    \label{fig:2Dplane}
\end{minipage}\qquad
\begin{minipage}{0.47\linewidth}
\centering
	\caption{Systematic uncertainties contributing to the \RDall\ results, with values in percent.}
	\vspace{0.2cm}
	\small
	\begin{tabular}{lcc}
	    \hline
		Source 								    &  $\Delta \RD$    &  $\Delta \RDSt$ \\
		\hline
		$D^{**}$ composition                    &                     0.76 &                      1.41 \\
		Fake $D^{(*)}$ calibration              &                     0.19 &                      0.11 \\
		$\btag$ calibration                     &                     0.07 &                      0.05 \\
		Feed-down factors                       &                     1.69 &                      0.44 \\
		Efficiency factors                      &                     1.93 &                      4.12 \\
		Lepton efficiency                       &                     0.36 &                      0.33 \\
		and fake rate &&\\
		Slow pion efficiency                    &                     0.08 &                      0.08 \\
		PDF shapes                              &                     4.39 &                      2.25 \\
		$B$ decay form factors                  &                     0.55 &                      0.28 \\
		Luminosity and $\mathcal{B}(\Upsilon(4S))$&                     0.10 &                      0.04 \\
		$\mathcal{B}(B \to D^{(*)} \ell \nu)$   &                     0.05 &                      0.02 \\
		$\mathcal{B}(D)$                        &                     0.35 &                      0.13 \\
		$\mathcal{B}(D^*)$                      &                     0.04 &                      0.02 \\
		$\mathcal{B}(\tauDec)$ 				    &                     0.15 &                      0.14 \\
		\hline
		Total                                	&                     5.21 &                      4.94 \\
		\hline
	\end{tabular}
	\label{tab:sys}
\end{minipage}
\end{table}

\normalsize

\section*{Acknowledgments}

We thank the KEKB group for excellent operation of the
accelerator; the KEK cryogenics group for efficient solenoid
operations; and the KEK computer group, the NII, and 
PNNL/EMSL for valuable computing and SINET5 network support.  
We acknowledge support from MEXT, JSPS and Nagoya's TLPRC (Japan);
ARC (Australia); FWF (Austria); NSFC and CCEPP (China); 
MSMT (Czechia); CZF, DFG, EXC153, and VS (Germany);
DST (India); INFN (Italy); 
MOE, MSIP, NRF, RSRI, FLRFAS project and GSDC of KISTI and KREONET/GLORIAD (Korea);
MNiSW and NCN (Poland); MSHE (Russia); ARRS (Slovenia);
IKERBASQUE (Spain); 
SNSF (Switzerland); MOE and MOST (Taiwan); and DOE and NSF (USA).
We acknowledge the support provided by the Albert Shimmins Fund for the writing of this proceedings.


\section*{References}

\bibliography{bibliography}

\begin{thebibliography}{10}

\bibitem{note}
{{Throughout this proceedings, the inclusion of the charge-conjugate decay mode
  is implied}}.

\bibitem{Bigi:2016mdz}
D.~Bigi and P.~Gambino.
\newblock {Revisiting $B\to D \ell \nu$}.
\newblock {\em Phys. Rev.}, D94(9):094008, 2016.

\bibitem{Bernlochner:2017jka}
F.~U. Bernlochner, Z.~Ligeti, M.~Papucci, and D.~J. Robinson.
\newblock {Combined analysis of semileptonic $B$ decays to $D$ and $D^*$:
  $R(D^{(*)})$, $|V_{cb}|$, and new physics}.
\newblock {\em Phys. Rev.}, D95(11):115008, 2017.
\newblock [erratum: Phys. Rev.D97, no.5, 059902 (2018)].

\bibitem{HFLAV}
Y.~Amhis et~al.
\newblock {Averages of $b$-hadron, $c$-hadron, and $\tau$-lepton properties as
  of summer 2016}.
\newblock {\em Eur. Phys. J.}, C77:895, 2017.
\newblock {updated results and plots available at
  \href{https://hflav.web.cern.ch}{{\texttt{https://hflav.web.cern.ch}}}}.

\bibitem{Sato:2016svk}
Y.~Sato et~al.
\newblock {Measurement of the branching ratio of $\bar{B}^0 \rightarrow D^{*+}
  \tau^- \bar{\nu}_{\tau}$ relative to $\bar{B}^0 \rightarrow D^{*+} \ell^-
  \bar{\nu}_{\ell}$ decays with a semileptonic tagging method}.
\newblock {\em Phys. Rev.}, D94(7):072007, 2016.

\bibitem{Abashian:2000cg}
A.~Abashian et~al.
\newblock {The Belle Detector}.
\newblock {\em Nucl. Instrum. Methods}, A479:117--232, 2002.
\newblock also see detector section in J.Brodzicka {\it et al.}, Prog. Theor.
  Exp. Phys. {\bf 2012}, 04D001 (2012).

\bibitem{KUROKAWA20031}
S.~Kurokawa and E.~Kikutani.
\newblock Overview of the {KEKB} accelerators.
\newblock {\em Nucl. Instrum. Methods}, A499(1):1 -- 7, 2003.
\newblock and other papers included in this Volume; T.Abe {\it et al.}, Prog.
  Theor. Exp. Phys. {\bf 2013}, 03A001 (2013) and references therein.

\bibitem{LANGE2001152}
D.~J. Lange.
\newblock The {E}vt{G}en particle decay simulation package.
\newblock {\em Nucl. Instrum. Methods Phys. Res., Sect. A}, 462(1):152 -- 155,
  2001.

\bibitem{Brun:1987ma}
R.~Brun et~al.
\newblock {GEANT3}.
\newblock {1987}.
\newblock {CERN Report No. DD/EE/84-1}.

\bibitem{Keck2019}
T.~Keck et~al.
\newblock The {F}ull {E}vent {I}nterpretation.
\newblock {\em Computing and Software for Big Science}, 3(1):6, 2019.

\bibitem{PDG}
M.~Tanabashi et~al.
\newblock Review of {P}article {P}hysics.
\newblock {\em Phys. Rev.}, D98:030001, 2018.

\bibitem{xgboost16}
T.~{Chen} and C.~{Guestrin}.
\newblock {XGBoost: A Scalable Tree Boosting System}.
\newblock 2016.
\newblock {1603.02754}.

\bibitem{huschle2015}
M.~Huschle et~al.
\newblock {Measurement of the branching ratio of $\bar{B} \to D^{(\ast)} \tau^-
  \bar{\nu}_\tau$ relative to $\bar{B} \to D^{(\ast)} \ell^- \bar{\nu}_\ell$
  decays with hadronic tagging at Belle}.
\newblock {\em Phys. Rev.}, D92(7):072014, 2015.

\end{thebibliography}

\end{document}